\documentclass[a4paper,11pt]{article}
\usepackage[left= 2.5cm,right = 2cm,headheight=0pt,top=2.5cm]{geometry}

\usepackage[T1]{fontenc} % Use 8-bit encoding that has 256 glyphs
\usepackage[english]{babel} % English language/hyphenation
\usepackage{amsmath,amsfonts,amsthm,bm} % Math packages

\DeclareMathOperator*{\argmin}{arg\,min}

\usepackage{graphicx,float}

\usepackage{setspace}
\usepackage[backend=bibtex,style=numeric,sorting=none,maxbibnames=5]{biblatex}
\bibliography{Library}

\usepackage{sectsty} % Allows customizing section commands
\usepackage{fancyhdr} % Custom headers and footers
\usepackage[ddmmyyyy]{datetime}

\begin{document}

%\begin{titlepage}
\title{On the relationship between a Gamma distributed precision parameter and the associated standard deviation in the context of Bayesian parameter inference}
\author{Manuel M. Eichenlaub\footnote{email: m.eichenlaub@warwick.ac.uk}}
\date{\small	
 	School of Engineering, University of Warwick, United Kingdom}
\maketitle
\begin{abstract}
\noindent 
In Bayesian inference, an unknown measurement uncertainty is often quantified in terms of a Gamma distributed precision parameter, which is impractical when prior information on the standard deviation of the measurement uncertainty shall be utilised during inference. This paper thus introduces a method for transforming between a gamma distributed precision parameter and the distribution of the associated standard deviation. The proposed method is based on numerical optimisation and shows adequate results for a wide range of scenarios.\\
\vspace{0in}
\vspace{0in}\\
\end{abstract}

\onehalfspacing
\section{Introduction} \label{sec:introduction}

In the context of Bayesian parameter inference, it is common to model the error associated with the observed data as follows \cite{daunizeau09}:

\begin{equation}
y(t) = g(\cdot)+\varepsilon \quad \text{with} \quad \varepsilon \sim \mathcal{N}(0,p^{-1}),
\end{equation}

where $y(t)$ is the observed data, $g(\cdot)$ the observation function and $\varepsilon$ the Gaussian distributed measurement error with zero mean and precision $p$. This precision is commonly described as being Gamma distributed with shape and rate parameters $a$ and $b$, respectively 

\begin{equation}
p \sim \mathcal{G}a(a,b) \quad \text{for} \quad p,a,b>0
\end{equation}

The values of $a$ and $b$ define the probability density function (PDF) over $p$ and are updated during parameter inference from a prior distribution of $p$, defined by $a_{0}$ and $b_{0}$. The use of a Gamma distribution over $p$ is justified by the fact that like the precision, the Gamma distribution is defined over positive values only. It furthermore forms a conjugate prior to the Gaussian distributed likelihood, therefore leading to analytically tractable posterior distributions and update rules \cite{daunizeau09,bishop06}. An example of this approach can be found in a variational Bayesian method for the identification stochastic nonlinear models \cite{daunizeau09,daunizeau14}.\\

The prior for $p$ is often chosen to be weak and uninformative \cite{gelman06}. However, in a number of practical applications, the collection of data is a known process and information on the measurement error $\varepsilon$ can be found in the literature. Here, the measurement error is commonly quantified in the form of the standard deviation $s$. An example for this would be the coefficient of variation of a certain immunoassay, e.g. of insulin. This information on $s$ can therefore be used to specify the prior PDF over $p$. Additionally, it is useful to allow the interpretation of the posterior distribution over $p$ in terms of $s$. This paper therefore introduces a method for the forwards and backwards transformation between the PDFs over $p$ and $s$.

\section{Methodology}

\subsection{Transformation from $p$ to $s$} 

The Gamma distribution over $p$ is defined by the following PDF of shape and rate parameters $a$ and $b$, respectively \cite{florescu14},

\begin{equation}
f_{p}(p|a,b) = \dfrac{b^{a}}{\Gamma(a)} p^{a-1} \exp(-pb) \quad \text{for} \quad p,a,b>0,
\end{equation}

where $\Gamma(\cdot)$ is the Gamma function. The mean and variance of this Gamma PDF are given by \cite{florescu14}
%---------------------------------------------------------------------------
\begin{equation} \label{eqn2:gamma_ss}
	\mathbb{E}[p]_{f_{p}} = \frac{a}{b} \quad \text{and} \quad \text{Var}[p]_{f_{p}} = \frac{a}{b^{2}}.
\end{equation}

The standard deviation $s$ of the measurement error $\varepsilon$ and its precision $p$ are related as follows:

\begin{equation}\label{rel_s_p}
s = \dfrac{1}{\sqrt{p}} 
\end{equation}

In order to determine the PDF of $s$ in terms of $a$ and $b$ the following theorem is used. If $f_{x}$ is a PDF over the random variable $x$ and the mapping $y = h(x)$ is introduced, then the PDF over the random variable $y$ is given by \cite{florescu14}:

\begin{equation}\label{inv}
f_{y}(y) = f_{x}(h^{-1}(y))\left\lvert \frac{\text{d}h^{-1}(y)}{\text{d}y}\right\rvert
\end{equation}

Defining $s=h(p)=1/\sqrt{p}$ from expression (\ref{rel_s_p}) and therefore $h^{-1}(s)=1/s^{2}$, expression (\ref{inv}) can be used to determine the PDF $f_{s}$ over $s$ as follows:

\begin{equation}
\begin{aligned}
f_{s}(s|a,b) &= f_{p}(\frac{1}{s^{2}}|a,b) \left\lvert \frac{\text{d}}{\text{d}s}\frac{1}{s^{2}}\right\rvert\\
& = \dfrac{b^{a}}{\Gamma(a)} \left(\dfrac{1}{s^{2}}\right)^{a-1} \exp\left(-\dfrac{b}{s^{2}}\right) \dfrac{2}{s^{3}} \\
& = \dfrac{2b^{a}}{\Gamma(a)} s^{-2a-1} \exp\left(-\dfrac{b}{s^{2}}\right) \\
\end{aligned}
\end{equation}

Using symbolic computation, this new probability distribution can be characterized by the following expression for the mean $\mu_{s}$ and the standard deviation $\sigma_{s}$, valid for $a>1$:

\begin{equation} \label{sig}
\begin{aligned} 
	\mu_{s} &= \mathbb{E}[s]_{f_{s}} = \sqrt{b}\dfrac{\Gamma(a-\frac{1}{2})}{\Gamma(a)},\\[12pt]
	\sigma_{s}^{2} &= \text{Var}[s]_{f_{s}} =  b\left[\dfrac{1}{a-1}-\dfrac{\Gamma(a-\frac{1}{2})^{2}}{\Gamma(a)^{2}}\right].
\end{aligned}
\end{equation}

To facilitate the numerical calculation, the logarithm of the Gamma function $\log\Gamma(\cdot)$ is used instead of the fast growing Gamma function itself. This modifies expressions (\ref{sig}) to give 

\begin{equation} \label{ln_sig}
	\begin{aligned}
		\mu_{s} &= \sqrt{b}\exp\left[\log\Gamma(a-\dfrac{1}{2})-\log\Gamma(a)\right] \\[12pt]
		\sigma_{s}^{2} &= b\left[\dfrac{1}{a-1}-\exp\left[\log\Gamma(a-\dfrac{1}{2})^{2}-\log\Gamma(a)^{2}\right]\right]	
	\end{aligned}	
\end{equation}

\begin{figure}[h]	
	\centering
	\includegraphics[scale=1]{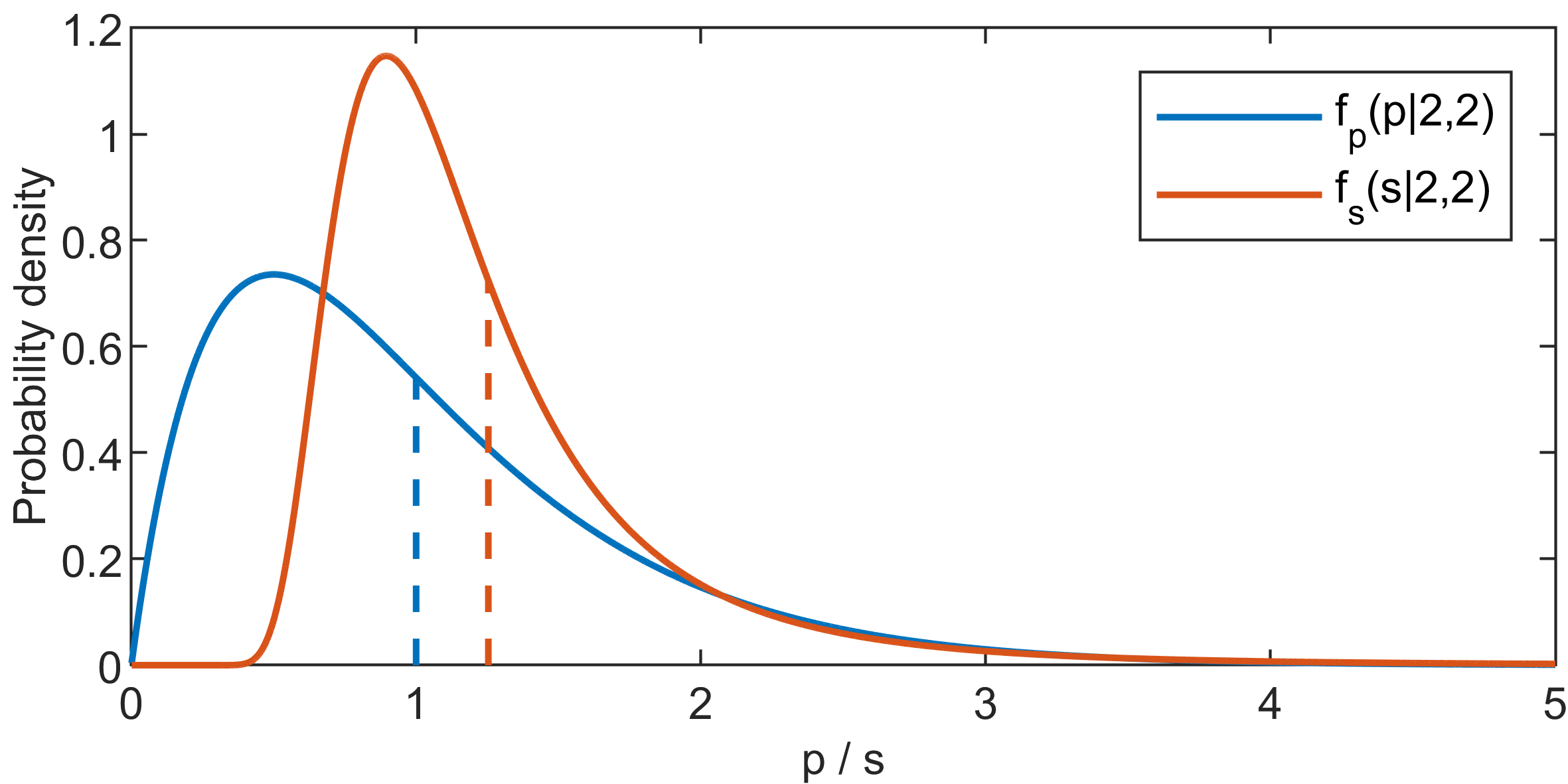}
	\caption[Example plots of $f_{p}$ and $f_{s}$]{Examples of the two PDFs of $f_{p}$ and $f_{s}$ for $a=b=2$. The dashed vertical lines display the values of the respective means.} \label{fig2:pdfs}
\end{figure}

An example of the PDFs over $p$ and $s$ is provided in Figure \ref{fig2:pdfs}, where it is demonstrated that the mean of $f_{p}$ does not simply transform into the mean of $f_{s}$ by applying the mapping $s=1/\sqrt{p}$. \\
%--------------------------------------------------------------------- 
%--------------------------------------------------------------------- 

\subsection{Transformation from $s$ to $p$} 

\begin{figure}[h]
	\centering
	\includegraphics[scale=1]{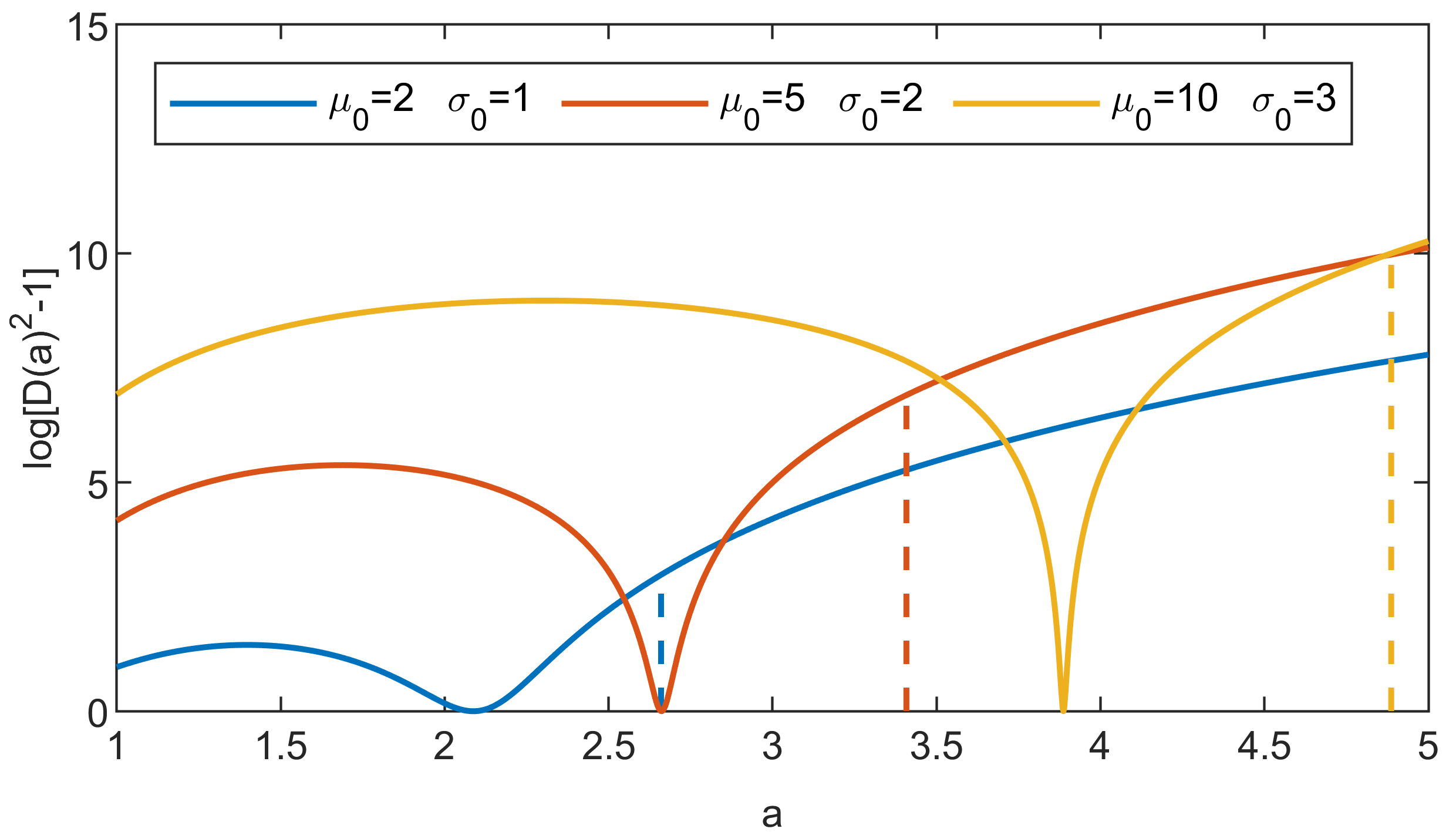}
	\caption[Example plots of $\log(D(a)^{2}+1)$]{Examples of the objective function $\log[D(a)^{2}+1]$ for differing values of of $\mu_{0}$ and $\sigma_{0}$. The minima correspond to the values of $a_{0}$, where $D(a_{0}) = 0$. The dashed lines give the respective values of the upper bound estimation of $\hat{a}_{0}$.} \label{fig2:logD}
\end{figure}

Expressions (\ref{ln_sig}) allows the interpretation of the posterior PDF of $p$, specified by $a$ and $b$ using the corresponding distribution over $s$. To specify the prior distribution over $p$ based on a chosen prior PDF over $s$, which in turn can be based on existing information, the following procedure is introduced. Defining $\mu_{0}$ and $\sigma_{0}$ characterising the the prior PDF over $s$, the goal is to calculate the associated values for $a_{0}$ and $b_{0}$, characterising the prior PDF over $p$. First, the following substitution is defined: 
%--------------------------------------------------------------------- 
\begin{equation} \label{eqn2:noi_S}
	S(a) = \dfrac{\Gamma(a-\frac{1}{2})^{2}}{\Gamma(a)^{2}} = \exp\left[\log\Gamma(a-\frac{1}{2})^{2}-\log\Gamma(a)^{2}\right].
\end{equation}
%--------------------------------------------------------------------- 
This is followed by the combination and reformulation of the expressions (\ref{ln_sig}) into 
%--------------------------------------------------------------------- 
\begin{equation}\label{eqn2:noi_D}
	D(a) = \dfrac{{\mu_{0}}^2}{S(a)} - \dfrac{{\sigma_{0}}^2}{\dfrac{1}{a-1}-S(a)} = 0.
\end{equation}
%--------------------------------------------------------------------- 
This eliminates $b$ and makes it possible to find $a_{0}$ by solving the equation $D(a_{0})=0$ and subsequently calculating $b_{0}$ using:
%--------------------------------------------------------------------- 
\begin{equation}\label{eqn2:noi_b0}
	b_{0} = \dfrac{{\mu_{0}}^2}{S(a_{0})}.
\end{equation}
%--------------------------------------------------------------------- 
To find $a_0$, expression (\ref{eqn2:noi_D}) is reformulated into a constrained numerical minimisation task:
%--------------------------------------------------------------------- 
\begin{equation}
	a_{0} = \argmin_{a} \left( \log\left[D(a)^{2}+1\right] \right) \quad \text{for} \quad a>1.
\end{equation}
%--------------------------------------------------------------------- 
The square operation and addition of one within the logarithm ensures that the objective function is always positive except for $\log[D(a_{0})^{2}+1]$, where the expression is zero. The logarithm facilitates the numerical calculations as the values of only $D(a)^{2}$ would grow rapidly as $a$ increases. An example of the objective function for different values of $\mu_{0}$ and $\sigma_{0}$ is given in Figure \ref{fig2:logD}, demonstrating a clear minimum of the objective function at $a_0$.\\

%--------------------------------------------------------------------- 
%--------------------------------------------------------------------- 

This minimisation can be numerically accomplished using an appropriate constrained implementation of an optimisation technique, e.g. \textit{fminbnd} in MATLAB or \textit{minimize\char`_scalar} in Scipy. Since the expression (\ref{ln_sig}) is only valid for $a>1$ the lower bound is set to 1. To specify the upper bound based on the given values of $\mu_{0}$ and $\sigma_{0}$, the function $S(a)$ from expression (\ref{eqn2:noi_S}) is approximated with the first two terms of its series expansion for $a\rightarrow \infty$, giving:
%---------------------------------------------------------------------
\begin{equation}\label{S_a}
	\hat{S}(a) = \dfrac{1}{a}+\dfrac{3}{4{a}^{2}}.
\end{equation}
%---------------------------------------------------------------------
With this approximation, the minimum $\hat{a}_{0}$ of $D(a)$ can be found analytically
%---------------------------------------------------------------------
\begin{equation} \label{ahat}
	\hat{a}_{0} = \dfrac{1}{8}\left[1+\sqrt{49+\dfrac{{\mu_{0}}^{4}}{{\sigma_{0}}^{4}}+50\dfrac{{\mu_{0}}^{2}}{{\sigma_{0}}^{2}}}+\dfrac{{\mu_{0}}^{2}}{{\sigma_{0}}^{2}}\right],
\end{equation}
%---------------------------------------------------------------------
which is used as the upper bound for the constrained minimisation (see Figure \ref{fig2:logD}). Expressions (\ref{S_a}) - (\ref{ahat}) were found using symbolic calculation. This procedure now allows the calculation of $a_{0}$ and subsequently $b_{0}$ with expression (\ref{eqn2:noi_b0}) based on $\mu_{0}$ and $\sigma_{0}$, thereby specifying the prior distribution over $p$. All symbolic calculations were done using Mathematica and the respective code is available online (https://github.com/manueich/Noise\_Gamma).  

\subsection{Validation}

In order to assess the accuracy of the numerical calculations when transforming from $s$ to $p$, the method was implemented in MATLAB 2020a with the function \textit{fminbnd} and Python with the function \textit{minimize\char`_scalar} using identical optimisation settings. The respective code is available online (https://github.com/manueich/Noise\_Gamma).\\

As a test, a wide range of possible values for $\mu$ and $\sigma$ are transformed into values for $a$ and $b$ and subsequently back-transformed into $\mu$ and $\sigma$ using expressions (\ref{ln_sig}). These results can then be compared to the starting values of $\mu$ and $\sigma$. A test is considered as passed if the original values for $\mu$ and $\sigma$ can be recovered with a relative error smaller than 1 \% on both parameters. For $\mu$, a total of 1000 values logarithmically scaled on a range between $10^{-4}$ and $10^{4}$ are chosen. Values outside this range should not be encountered in practice. Subsequently, for each value of $\mu$, a total of 1000 values for $\sigma$ are logarithmically scaled on a range between $10^{-4} \cdot \mu$ and $10^{2} \cdot \mu$ are tested.

\section{Results and discussion}

\begin{figure}[h]
	\centering
	\includegraphics[scale=1]{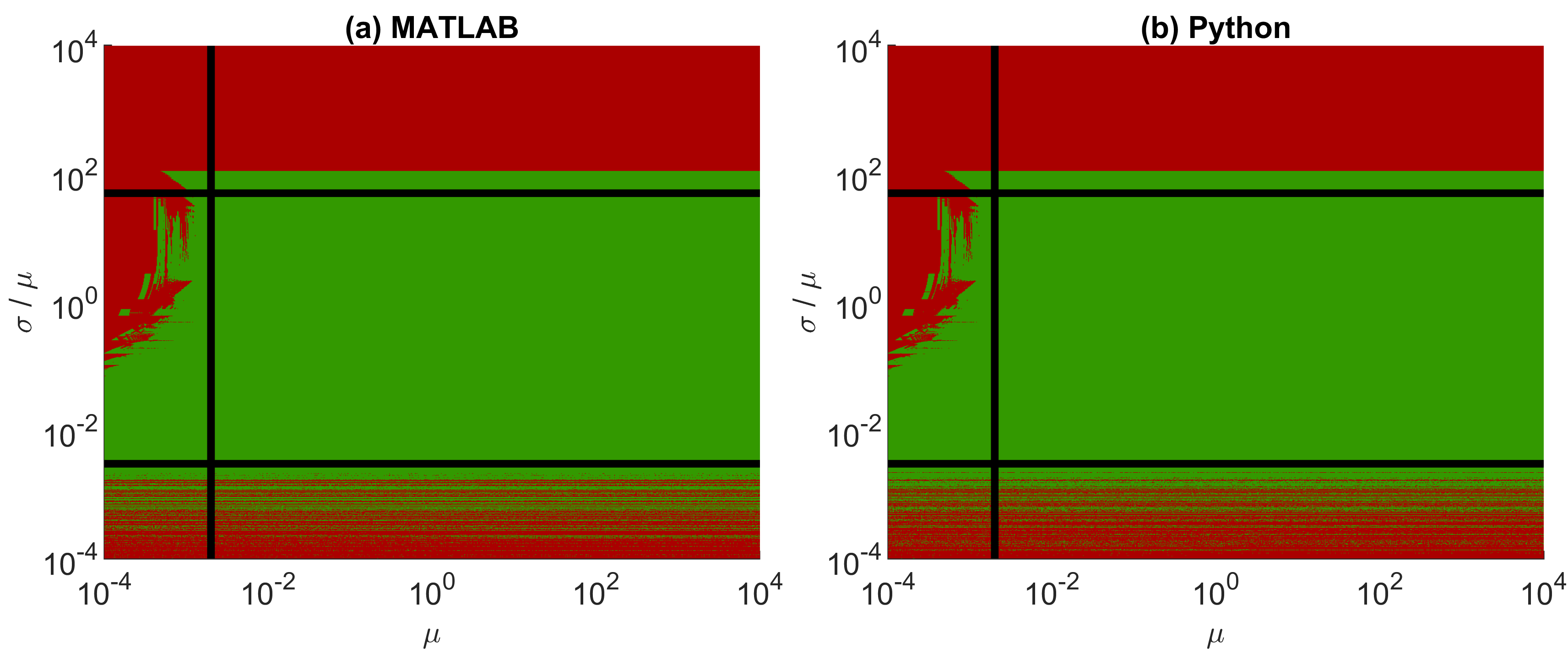}\\
	\caption{Results of the validation procedure for (a) MATLAB and (b) Python. Red indicates a failed and green indicates a passed test. The solid black lines give the suggested cut-off values.} \label{fig_res}
\end{figure}

The results of the validation procedure for both MATLAB and Python are displayed in Figure \ref{fig_res} and are very similar. Based on this the following cut-off values are proposed, which should ensure a robust calculation of $a$ and $b$ based on $\mu$ and $\sigma$.
\begin{itemize}
	\item $2 \cdot 10^{-3} < \mu < 10^{4}$\\
	\item $3 \cdot 10^{-3} < \dfrac{\sigma}{\mu} < 50$\\
\end{itemize}

These ranges should cover a large number of possible values that could be encountered in practice.
  
%--------------------------------------------------------------------- 
%--------------------------------------------------------------------- 

\vspace{1cm}

\printbibliography

\end{document}